\documentclass[superscriptaddress,aps,prb,twocolumn,amsmath,amssymb]{revtex4-1}
\usepackage{graphicx}
\usepackage{dcolumn}
\usepackage{color,soul} 
\usepackage{multirow}
\usepackage{longtable}
\usepackage{float}
\usepackage[colorlinks=true, citecolor=red, urlcolor=blue]{hyperref}
\usepackage{threeparttable}

\usepackage{color,ulem}

\begin{document}

\title{A Computational Framework for Automation of Point Defect Calculations}

\author{Anuj Goyal}
\email{Anuj.Goyal@nrel.gov}
\affiliation{Colorado School of Mines, Golden, CO 80401, USA}
\affiliation{National Renewable Energy Laboratory, Golden CO 80401, USA}
\author{Prashun Gorai}
\affiliation{Colorado School of Mines, Golden, CO 80401, USA}
\affiliation{National Renewable Energy Laboratory, Golden CO 80401, USA}
\author{Haowei Peng}
\affiliation{National Renewable Energy Laboratory, Golden CO 80401, USA}
\author{Stephan Lany}
\affiliation{National Renewable Energy Laboratory, Golden CO 80401, USA}
\author{Vladan Stevanovi\'{c}}
\email{Vladan.Stevanovic@nrel.gov}
\affiliation{Colorado School of Mines, Golden, CO 80401, USA}
\affiliation{National Renewable Energy Laboratory, Golden CO 80401, USA}


\begin{abstract}
A complete and rigorously validated open-source Python framework to automate point defect calculations using density functional theory has been developed. The framework provides an effective and efficient method for defect structure generation, and creation of simple yet customizable workflows to analyze defect calculations. The package provides the capability to compute widely-accepted correction schemes to overcome finite-size effects, including (1) potential alignment, (2) image-charge correction, and (3) band filling correction to shallow defects. Using Si, ZnO and In$_{2}$O$_{3}$ as test examples, we demonstrate the package capabilities and validate the methodology.
\end{abstract}
\maketitle
\section{\label{sec:I}Introduction}

\begin{figure}[t!]
\includegraphics[width=0.9\linewidth]{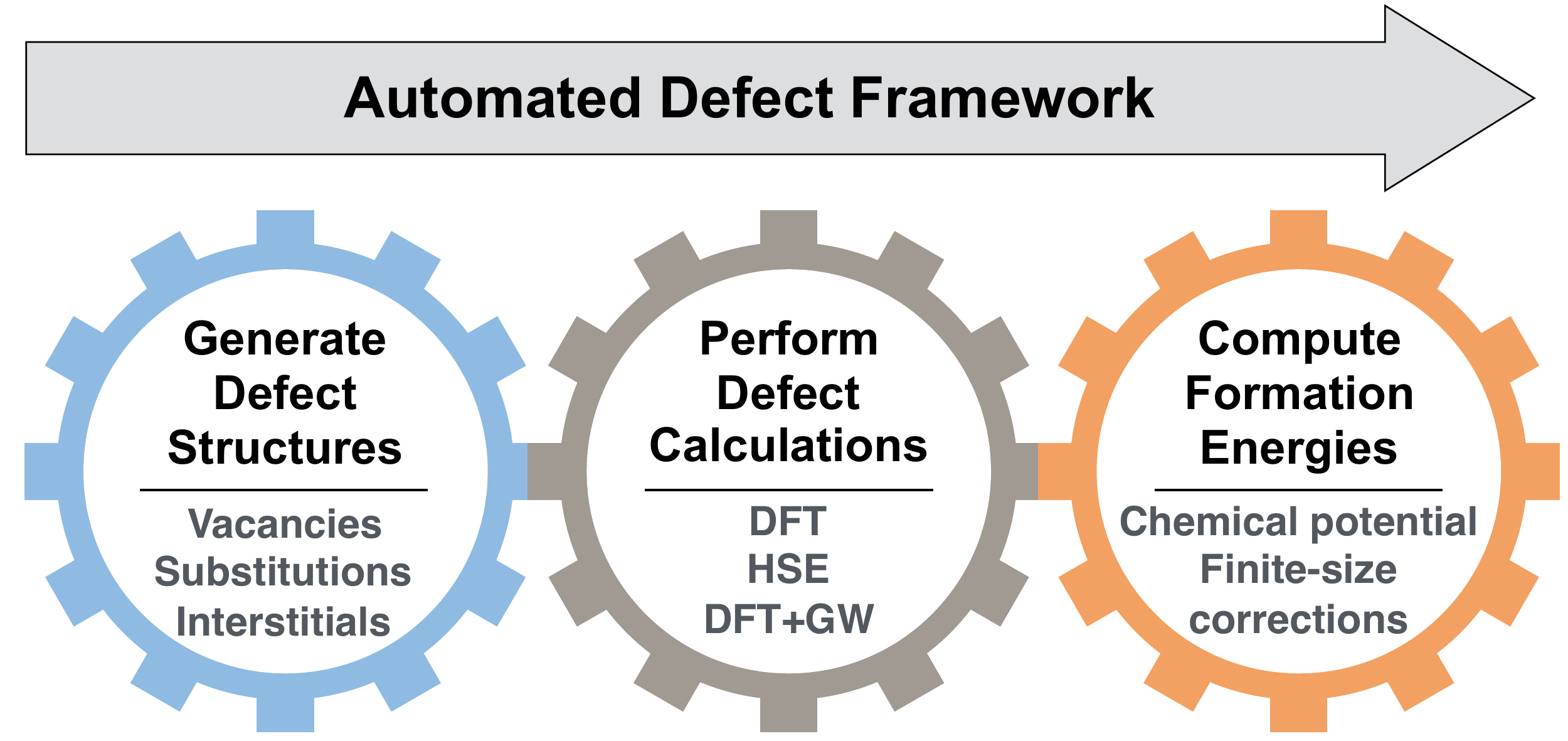}
\caption{\label{fig:1}Three key components of the computational framework to automate point defect calculations.}
\end{figure}

In semiconductor materials, point defects play a vital role in determining their properties and performance, particularly in microelectronics,\cite{DelAlamo2011} optoelectronics,\cite{Yu2016} and thermoelectrics\cite{Yan2015} related applications. The dominant point defects and their concentrations are determined from the defect formation energies, which can be predicted with reasonable accuracy\cite{Lejaeghere2016} using first-principles methods such as density functional theory (DFT). Therefore, computational modeling of point defects is increasingly becoming an indispensable tool to understand and predict behavior of semiconductors.\cite{VandeWalle2004, Alkauskas2011, Freysoldt2014} Modern approaches to point defect calculations uses DFT and are typically based on the supercell approach.\cite{Alkauskas2011, Freysoldt2014} With the goal of accelerating the design and discovery of materials by large-scale deployment of defect calculations, we have developed a computational framework (Fig. \ref{fig:1}) to automate supercell-based point defect calculations with DFT. Our approach successfully addresses two main challenges of automating point defect calculations: (1) generation of defects structures including vacancies, substitutional defects and interstitials, and (2) application of the finite-size and band gap corrections.\par

In the context of structure generation, creating supercells with vacancies and substitutional defects is relatively straightforward. In contrast, identifying likely locations of interstitials is much more challenging because of the large number of interstitialcy sites, especially in complex, multinary systems. In addition, interstitials might adopt complex configurations, including the split or dumbbell where the interstitial is associated with a off-site lattice atom. To address these challenges, we have developed an efficient scheme based on Voronoi tessellation;\cite{Rycroft2009} the scheme considers corners, edge and face centers of the Voronoi polyhedra as likely sites for interstitials. We demonstrate that, upon relaxing the structure, this scheme successfully discovers both the symmetric and general Wyckoff positions as well as the split interstitial configurations. Our implementation of this scheme is independent of pymatgen\cite{Ong2013} where Voronoi tessellation is also employed. Here we will discuss the algorithm in detail and validate the Voronoi-driven approach to identify interstitial sites.\par

Within the supercell approach to calculate the defect formation energies, finite-size artifacts need to be removed using carefully designed correction schemes. We have implemented tools to calculate the following finite-size corrections: (1) potential alignment, (2) image-charge correction, and (3) band filling correction to address Moss-Burstein-type effects. We follow the widely used and tested approach of Lany and Zunger\cite{Lany2008, Lany2009} out of the several others that addresses the same issues.\cite{Makov1995, Freysoldt2011, Taylor2011, Komsa2012, Kumagai2014} However, the automated framework is highly modular so that other correction schemes can be easily implemented including computation of defect formation energies using series of supercell sizes in order to extrapolate the values to the infinitely large supercell. In addition, the framework employs fitted elemental-phase reference energies (FERE) \cite{Lany2008b, Stevanovic2012} to compute elemental chemical potentials.
\begin{figure*}[t]
\includegraphics[width=0.75\textwidth]{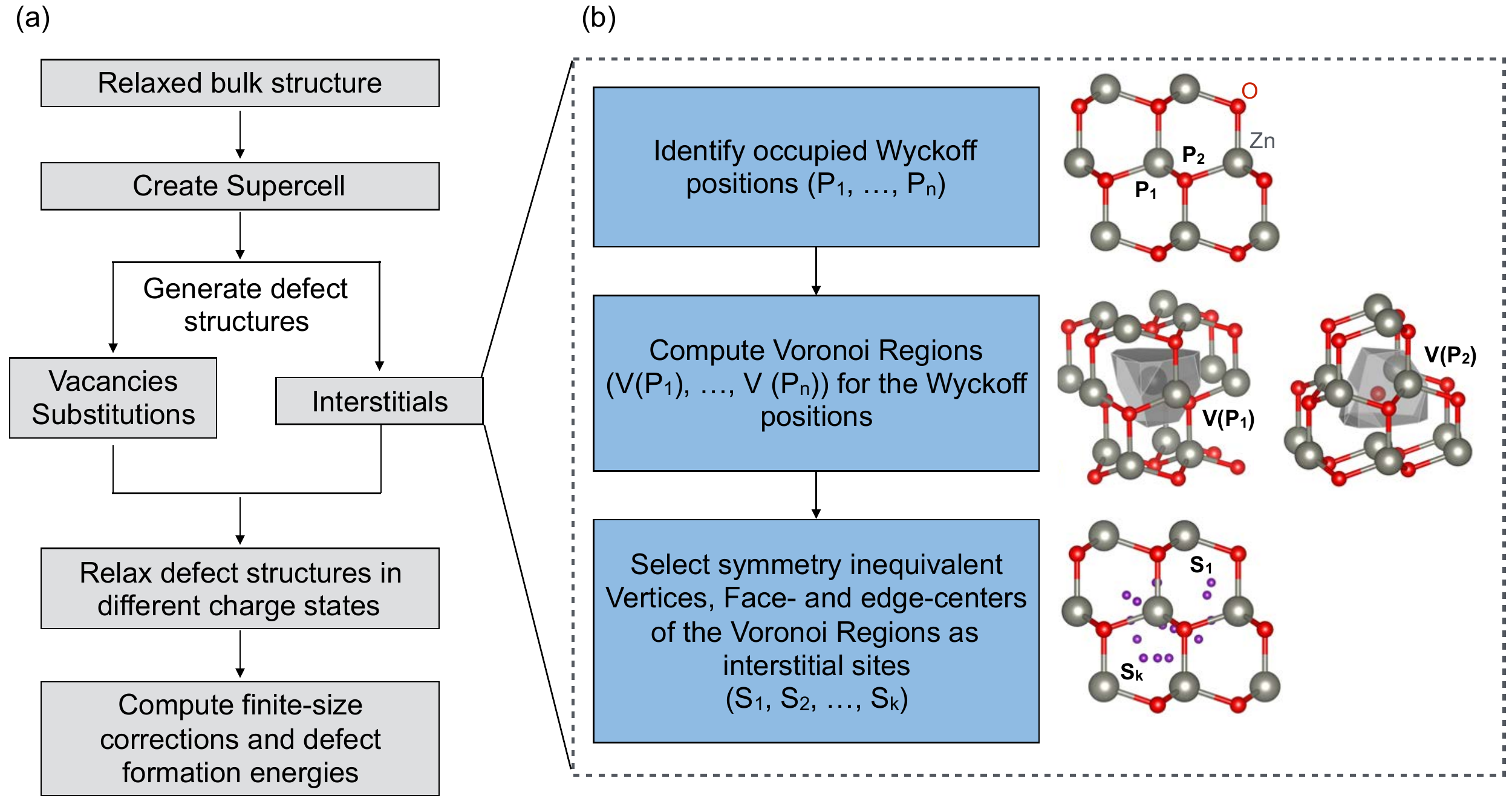}
\caption{\label{fig:2}(a) Workflow to perform defect calculations, and (b) the Voronoi tessellation-based algorithm to find interstitial sites in a given structure (example shown: ZnO).}
\end{figure*}

Beyond the finite-size effects, another source of inaccuracy arises from the well-known DFT band gap problem. Accurate band gaps are needed to correctly describe the formation energy of charged defects as a function of the electronic chemical potential i.e., Fermi energy. We employ state-of-the-art GW quasiparticle energy calculations\cite{Lars1965} to compute band edge shifts (relative to the DFT-computed band edges). The band edge shifts are used to correct the defect formation energy in multiple charge states. The automated framework is also capable of performing defect calculations with DFT hybrid functionals.\cite{Perdew1996a, Heyd2003} However, supercell defects calculations with hybrid functionals have sources of uncertainty arising from the choice of input parameters (e.g. fraction of exchange) and have considerable computational overheads.\cite{Peng2013} Therefore, we have implemented a DFT+GW approach for calculating defect formation energy that has been shown to be as accurate as calculations with hybrid functionals.\cite{Peng2013} Finally, we illustrate and validate the automated computational framework by considering the set of three well-studied semiconductor materials, Si, ZnO, and In$_{2}$O$_{3}$ with a total of 17 unique interstitial and vacancy structures in multiple charged states. We show that our results on defect formation energies and charge defect transition levels in Si, ZnO and In$_{2}$O$_{3}$ agree well with the literature. The framework successfully identifies the known intrinsic interstitial and vacancy structures in Si, ZnO and In$_{2}$O$_{3}$. In addition, it discovers interstitial structures in In$_{2}$O$_{3}$, with formation energies $\sim$0.5 eV above that of previously known interstitial structures.

\section{\label{sec:II} Overview of the Automated Defect Framework}

Figure \ref{fig:2}(a) presents a workflow of the automated framework, including generation of defect structures, relaxation of defect supercells within DFT using the PyLada framework,\cite{pylada} and determination of finite-size and band gap corrections to compute the defect formation energies. In this section, we describe each component of the framework and provided technical details.
\subsection{\label{sec:IIA} Generate Defect Structures}
The workflow takes the fully-relaxed primitive cell as an input to create supercells. To create a vacancy or substitutional defect in supercell, the occupied Wyckoff positions (lattice-sites) for all atom types in the supercell are identified. Then the corresponding atom is removed or substituted with an impurity atom, to generate vacancy or substitutional defect. Finally, the first nearest-neighbor atoms to the vacancy or substitutional site are randomly displaced ($\sim$ 0.1 \AA) to break the underlying site symmetry and thereby, ensuring the non-symmetric configurations of the defects are properly captured. The Voronoi tesselation,\cite{Rycroft2009, tess} scheme is employed to identify likely interstitial sites. Voronoi region is the volume that encloses the points $p$ closest to a given lattice site $P_{i}$ than to any other lattice site $P_{j}$ for $i,j \in I_{n} = \{1, ..., n\}$. Mathematically, it is defined as\cite{Rycroft2009}

\begin{equation}\label{eq:1}
V(P_{i}) = \{p \mid d(p, P_{i}) \leq d(p, P_{j}) \} \ \mathrm{for} \ j \neq i, j \in I_{n}
\end{equation}
where, $V(P_{i})$ is the Voronoi region associated with $P_{i}$, and $d(p, P_{i})$ specifies the minimum distance between a general point $p$ and $P_{i}$. To create an interstitial, Voronoi regions (Eq. \ref{eq:1}) are computed across each occupied Wyckoff positions, and symmetry inequivalent vertices, face, and edge centers of the Voronoi regions are chosen as the candidate sites for the interstitials. The number of candidate interstitial sites depends on the symmetry of the crystal structure. The lower the symmetry and the more complex the crystal structure, the larger the number of sites. For example, in In$_{2}$O$_{3}$ (space group Ia-3, 40 atoms in primitive cell), we find that some of the faces of the Voronoi region are very small, resulting in sampled interstitial sites very close to each other. Therefore a minimum tolerance of 0.5 \AA \ is used while determining symmetry inequivalent sites. The procedure is described in Fig. \ref{fig:2}(b), with ZnO as an example structure. 


\subsection{\label{sec:IIB} Perform Defect Calculations}

DFT calculations are performed with the projector augmented wave (PAW) method\cite{Blochl1994} as implement in VASP.\cite{Kresse1996a} The Perdew Burke Ernzerhof (PBE) exchange correlation functional\cite{Perdew1996} is used both in GGA (Si, In$_{2}$O$_{3}$) and GGA+U spin polarized calculations (ZnO, U(Zn-d) = 6 eV). The plane wave energy cutoff of 340 eV, and a Monkhorst-Pack k-point sampling\cite{Monkhorst1976} is used. The structures are taken from the inorganic crystal structure database (ICSD)\cite{Belsky2002} and fully relaxed using the procedure outlines in Ref.\citenum{Stevanovic2012}. Defect calculations are performed on 216, 96 and 80 atoms supercell for Si, ZnO, and In$_{2}$O$_{3}$, respectively.  A $\Gamma$-centered 2x2x2 k-point mesh is used for all supercell calculations, except for Si, for which only single $\Gamma$ point only calculations are performed. The low-frequency total (electronic + ionic) dielectric constant is obtained following the procedure in Ref.\citenum{Peng2013}. For hybrid functional (HSE06\cite{Krukau2006}) calculations in Si, the exchange-mixing, $\alpha=0.25$ is used. GW calculations are performed on the DFT relaxed structures, with the unit cell vectors re-scaled to match the experimental lattice volume.\cite{Peng2013} The high-throughput DFT calculations are performed using PyLada,\cite{pylada} a powerful Python framework for the constructing workflows, and managing large number of calculations. PyLada also offers variety of useful tools for constructing crystal structures, for manipulating functionals and extracting their output, and analyzing results.\cite{Gorai2016, Stevanovic2016, Deml2016}

\begin{figure}[t]
\includegraphics[width=0.8\linewidth]{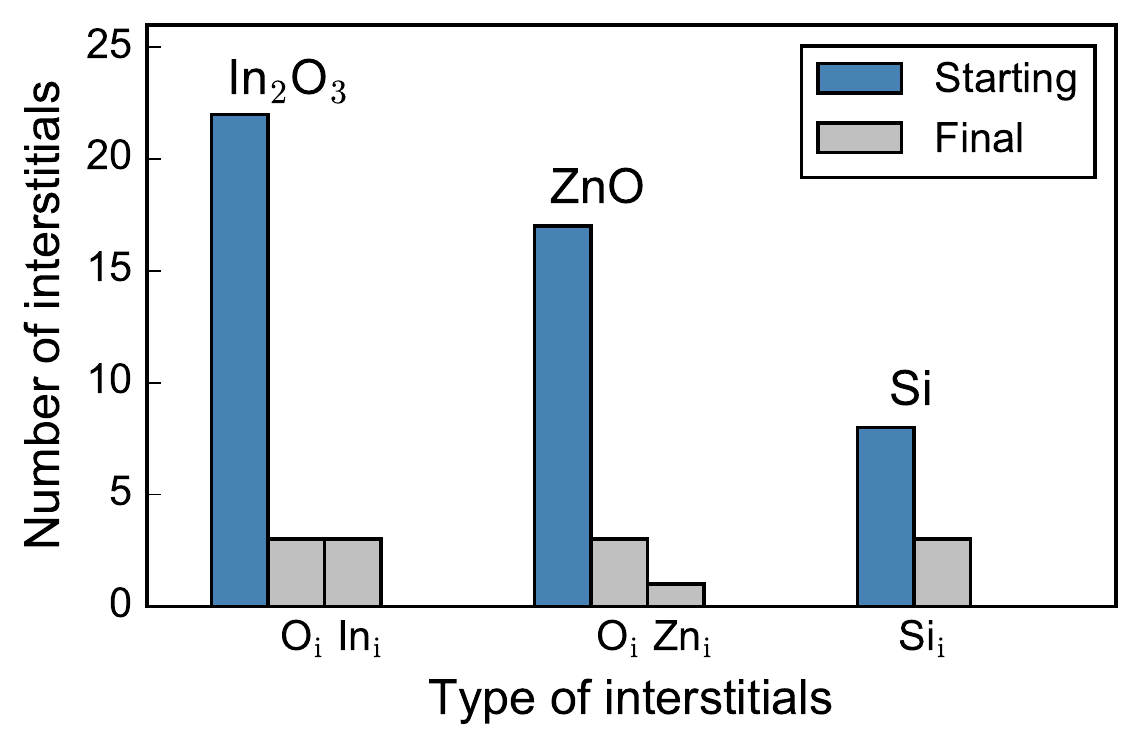}
\caption{\label{fig:3}Number of distinct starting (blue) and final DFT relaxed (silver) interstitial structures in Si, ZnO, and In$_{2}$O$_{3}$.}
\end{figure}

As summarized in Fig. \ref{fig:2}(a), the workflow starts with fully relaxing (volume, cell shape and ionic positions) the bulk primitive cell. Dielectric constant, and GW calculations are performed on the relaxed primitive cell. Point defects are then created in the bulk supercell followed by relaxation (only ionic positions) of defect structures in multiple charge states. Calculations of interstitial defects are performed in two steps: (1) All candidate interstitials (shown as starting interstitials in Fig. \ref{fig:3}) are relaxed in the neutral charge state, (2) followed by relaxation of only unique interstitials (shown as final interstitials in Fig. \ref{fig:3}) in multiple charge states. Unique interstitial structures are determined based on: (1) the total energy, (2) space group, and (3) the number of neighboring atoms. Finally, the defect formation energies are computed as discussed in the next section.\par

\begin{figure*}
\includegraphics[width=0.8\textwidth]{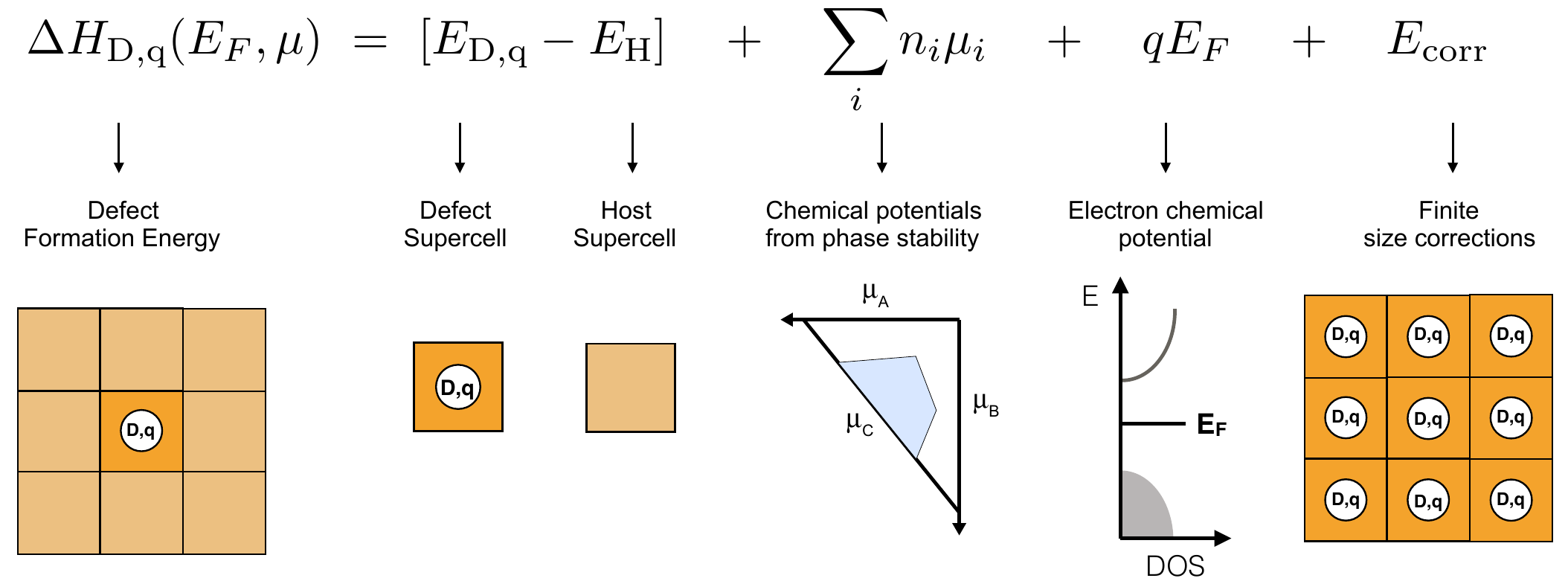}
\caption{\label{fig:4} Equation to compute charge defect formation energy as function of the chemical potential and Fermi energy.}
\end{figure*}


\subsection{\label{sec:IIC} Compute Defect Formation Energy}

The formation energy of the defect D in the charge state q is calculated as 
\begin{align}\label{eq:2}
\Delta H_{\mathrm{D, q}} (E_{F}, \mu) =  [E_{\mathrm{D,q}} - E_{\mathrm{H}}] + \sum_{i}n_{i} \mu_{i} + qE_{F} + E_{\mathrm{corr}}
\end{align}
 where, $E_{\mathrm{D,q}}$ and $E_{\mathrm{H}}$ are the total DFT energies of the defect and host supercell, respectively. $\mu_{i}$ is the chemical potential of the atom (host or impurity) of type $i$ added ($n_{i} < 0$) or removed ($n_{i} > 0$) from the host supercell to form the defect. $E_{F}$ is the Fermi energy, and $E_{\mathrm{corr}}$ is the term that account for the finite-size corrections, within the supercell approach. A schematic of Eq. \ref{eq:2}, representing computation of the defect formation energy from supercell to the dilute limit is shown in Fig. \ref{fig:4}.
 
\subsubsection{\label{sec:IIC1} Chemical Potential and Phase Stability}

Chemical potentials $\mu_{i}$ reflect the energy of the reservoirs for the atoms that are involved in creating the defect. Numerical values of the chemical potentials ($\mu_{i} = \mu_{i}^{0} + \Delta \mu_{i}$) depend on their implicit references, $\mu_{i}^{0}$, which here are obtained from the reference FERE \cite{Lany2008b, Stevanovic2012} energies, $\mu_{i}^{0} = \mu_{i}^{\mathrm{FERE}}$. FERE energies are also used to compute the formation enthalpy ($\Delta H_{f}$) of all the competing phases which are needed to determine the thermodynamic limits of the chemical potential $\Delta \mu_{i}$. The computed $\Delta H_{f}$ and $\mu_{i}^{\mathrm{FERE}}$ values are summarized in table \ref{tab:1}. For Si, $\mu_{Si}^{0} = E^{\mathrm{GGA}}$ (Si) = -5.41 eV/atom is used and to determine the limits to the respective elemental chemical potentials, we apply the following thermodynamic stability conditions, $\Delta \mu_{\mathrm{Zn}} + \Delta \mu_{\mathrm{O}} = \Delta H_{f}$(ZnO) and $ 2\Delta \mu_{\mathrm{In}} + 3\Delta \mu_{\mathrm{O}} = \Delta H_{f}$(In$_{2}$O$_{3}$), in ZnO and In$_{2}$O$_{3}$, respectively.

\subsubsection{\label{sec:IIC2} Electron Chemical Potential}

Fermi energy is the measure of the chemical potential of electrons. It is defined with respect to the host valence band maximum (VBM), $E_{F} = E_{\mathrm{VBM}}^{\mathrm{Host}} + \Delta E_{F}$, and is bounded by the conduction band maximum (CBM). DFT (GGA) band gaps are corrected by determining the band edge shifts, $\Delta E_{V}$ for the VBM, and $\Delta E_{C}$ for the CBM, from the GW quasiparticle energy calculations.\cite{Peng2013} The computed band gaps are also summarized in table \ref{tab:1}.

\begin{table*}[t]
	\caption{\label{tab:1} Calculated lattice parameters, dielectric constants (electronic, $\varepsilon_{\mathrm{elec.}}$, and ionic $\varepsilon_{\mathrm{ionic}}$), chemical potential, enthalpy of formation and band gap in Si (Fd-3m, 227), ZnO (P6$_{\mathrm{3}}$mc, 186) and In$_{2}$O$_{3}$ (Ia-3, 206). Experimental values are also cited.}
	\begin{ruledtabular}
		\begin{tabular}{ccccccc}
		 System & \multicolumn{1}{c}{Lattice constant} & \multicolumn{2}{c}{Dielectric constant} & \multicolumn{1}{c}{Chemical potential} & \multicolumn{1}{c}{$\Delta$H$_{f}$} & \multicolumn{1}{c}{Band Gap (eV)}  \\[0.4em]
		 & \multicolumn{1}{c}{(\AA)} & \multicolumn{1}{c}{$\varepsilon_{\mathrm{elec.}}$} & \multicolumn{1}{c}{$\varepsilon_{\mathrm{ionic}}$} & \multicolumn{1}{c}{$\mu_{i}^{\mathrm{FERE}}$ (eV)} & \multicolumn{1}{c}{(eV)} & \multicolumn{1}{c}{GW (GGA)} \\[0.4em]
		\colrule \\ [-0.7em]
		Si													&5.46	&13.36	&	&-5.41	&	&1.29 (0.62)\\[0.2em]
		Expt.	\footnote{References\citenum{Hubbard1975, Madelung1991}}		&5.43	&11.7	&	&		&	&1.17\\
\hline \\[-0.7em]
		ZnO			&$a$ = 3.28, 		&5.53	&5.12	&O = -4.76, 	&-3.63	&3.25 (0.73)\\[0.2em]
					&$c$ = 5.30		&		&		&Zn = -0.56	&		&\\[0.2em]
		Expt.	\footnote{References\citenum{Karzel1996, Ashkenov2003, Kubaschewski1993, Ozgur2005}}		&$a$ = 3.25, $c$ = 5.2	&3.7 - 3.8	&4 - 5.13	&	&-3.62	&3.44\\
\hline \\[-0.7em]
		In$_{2}$O$_{3}$	&10.28	&4.90	&6.47	&O = -4.76	&-9.45	&2.47 (0.96)\\[0.2em]
						&		&		&		&In = -2.31	&		&\\[0.2em]
		Expt.	\footnote{References\citenum{DeWit1977, Hamberg1986, Feneberg2016, Kubaschewski1993, Bourlange2008, Walsh2008, King2009, Irmscher2014}}			&10.1		&4.08	&4.8		&	&-9.6	&2.67 - 3.1\\
		\end{tabular}
	\end{ruledtabular}
\end{table*}

\subsubsection{\label{sec:IIC3} Finite-size Corrections}

Finite size corrections are implemented in the package following the approach of Lany and Zunger.\cite{Lany2008, Lany2009} Correction schemes focusing on single physical effect are considered.\cite{Freysoldt2014} These include: 

\textit{Potential alignment correction}, which restores the relative position of the host VBM in the calculations of charged defect (affecting the Fermi energy). Correction to the defect formation energy due to the potential alignment is given as\cite{Lany2008}

\begin{equation}\label{eq:3}
E_{\mathrm{PA}} (\mathrm{D,q}) = \mathrm{q} (V^{r}_{\mathrm{D,q}} - V^{r}_{\mathrm{H}})
\end{equation}
where the reference potentials, $V^{r}$, for the charged defect (D,q) and the pure host (H) are determined from the (local) atomic-sphere-averaged electrostatic potentials at the atomic sites farther away from the defect.\cite{Lany2008}

\textit{Image-charge correction}, is needed to correct for the spurious electrostatic interactions of the charged defect (in the presence of homogeneous compensating background charge) with its periodic images. This is given as\cite{Lany2009}

\begin{equation}\label{eq:4}
E_{\mathrm{IC}} = \bigg[1 + c_{sh}\big(1-\frac{1}{\varepsilon}\big) \bigg] \frac{q^{2}\alpha_{M}}{2 \varepsilon L}
\end{equation}
where $L = \Omega^{-1/3}$ is the linear supercell dimension (volume, $\Omega$), $\varepsilon$ is the static dielectric constant (electronic + ionic), and $\alpha_{M}$, $c_{sh}$ are the Madelung constant, and shape factor, respectively, for the adopted supercell geometry.

\textit{Band filling correction}, correct for the Moss-Burstein-type band filling effects that appear due to high defect concentrations in a typical finite-size supercell calculations.\cite{Lany2008} For a given k-point set (weighted sum, $w_{k}$) and band occupations, $\eta_{n,k}$, the correction for the shallow donor is computed as\cite{Lany2008}

\begin{figure*}
\includegraphics[width=0.8\textwidth]{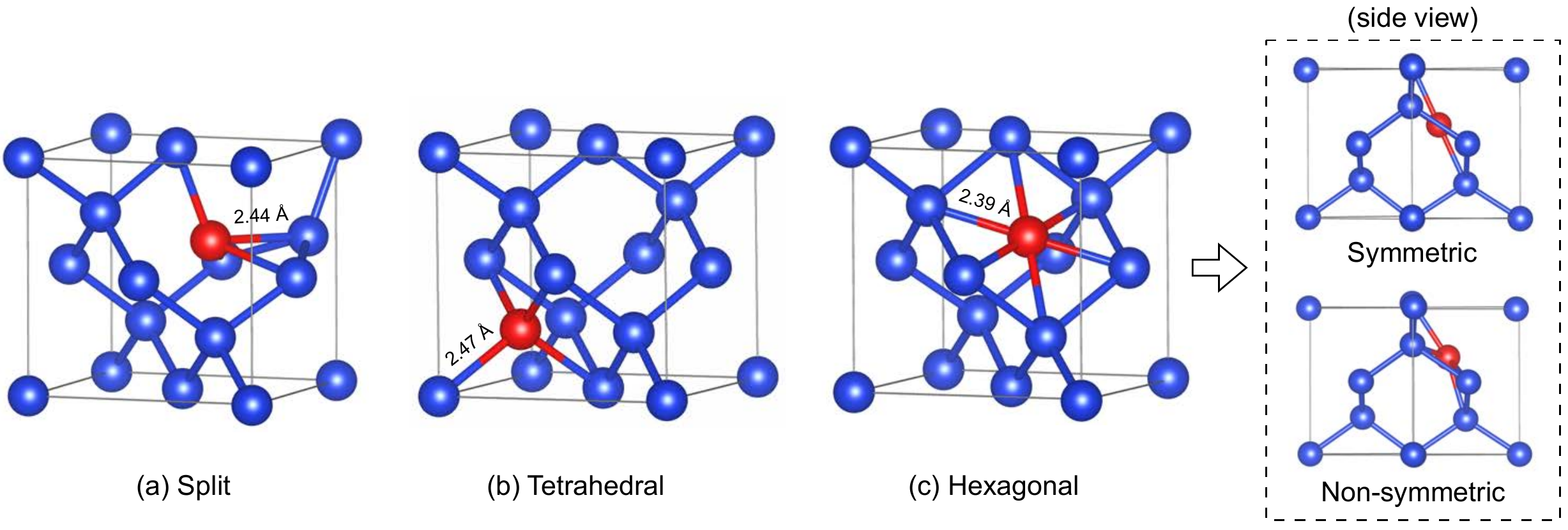}
\caption{\label{fig:5} Self interstitial structures in Si: (a) split, (b) tetrahedral, and (c) hexagonal geometry. Hexagonal interstitial has symmetric and non-symmetric 'C$_{v}$' configuration as shown in the projection along [110] direction in the side view.}
\end{figure*}

\begin{equation}\label{eq:5}
E_{\mathrm{BF}} = - \sum_{n,k} w_{k} \eta_{n,k}\big[e_{n.k} - \tilde{e_{C}}\big]
\end{equation}
and for shallow acceptor

\begin{equation}\label{eq:6}
E_{\mathrm{BF}} = \sum_{n,k} w_{k}(1- \eta_{n,k})\big[e_{n.k} - \tilde{e_{V}}\big]
\end{equation}
where, $e_{n.k}$ are the band energies in the defect calculation, $\tilde{e_{C}}$ is the CBM and $\tilde{e_{V}}$ is the VBM energy of the pure host after potential alignment correction.

\section{\label{sec:III}Examples}

\subsection{\label{sec:IIIA} Silicon}

Silicon has been the focus of both experimental\cite{Watkins1997, Bracht2007} and theoretical\cite{Baraff1979, Car1984, Bar-Yam1984, Puska1998, Wright2006, Rinke2009, Spiewak2013} research on intrinsic point defects over the past decade. Structure of both silicon vacancies\cite{Baraff1979, Puska1998, Wright2006, Spiewak2013} and self-interstitials\cite{Car1984, Bar-Yam1984, Rinke2009} has been topic of interest as they exists in several stable and metastable configurations.\par

In agreement with the existing literature, we find three distinct silicon self-interstitial structures namely, spit, hexagonal and tetrahedral as shown in Fig. \ref{fig:5}, among the starting 7 candidate sites from the defect generation code. Neutral split interstitial has the lowest formation energy (3.25 eV), followed by hexagonal interstitial with energy 0.2 eV higher. Hexagonal interstitial lies along the [111] direction and sit symmetrically at the center of the hexagon formed by six neighboring Si lattice atoms. Hexagonal interstitial also exists in a non-symmetric configuration, as shown in the side view in Fig. \ref{fig:5}. This configuration is 5 meV lower in energy than the symmetric one, but is unstable and relaxes to tetrahedral geometry in the charge states 1+ and 2+. Metastable hexagonal configuration have been reported in previous DFT calculations,\cite{Rinke2009} referred as `displaced hexagonal' or by C$_{3v}$ site symmetry. Tetrahedral interstitial in the neutral charge state has the highest formation energy, about 0.33 eV higher than the split interstitial. All the distances between the interstitial and the four neighboring Si lattice sites are same and are equal to 2.47 \AA.\par

\begin{figure}[b]
\includegraphics[width=1.\linewidth]{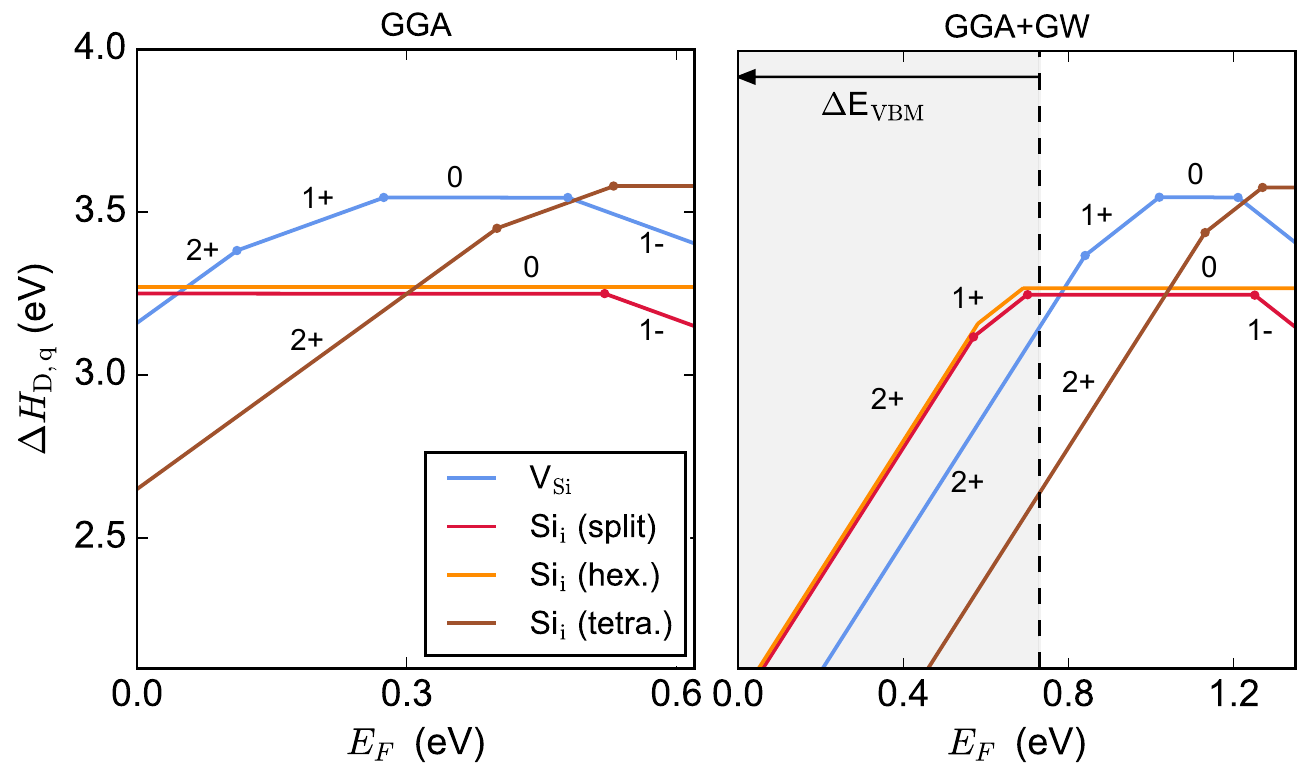}
\caption{\label{fig:6} Defect formation energy as function of the Fermi energy from GGA and GGA+GW calculations.}
\end{figure}

\begin{figure*}
\includegraphics[width=0.8\textwidth]{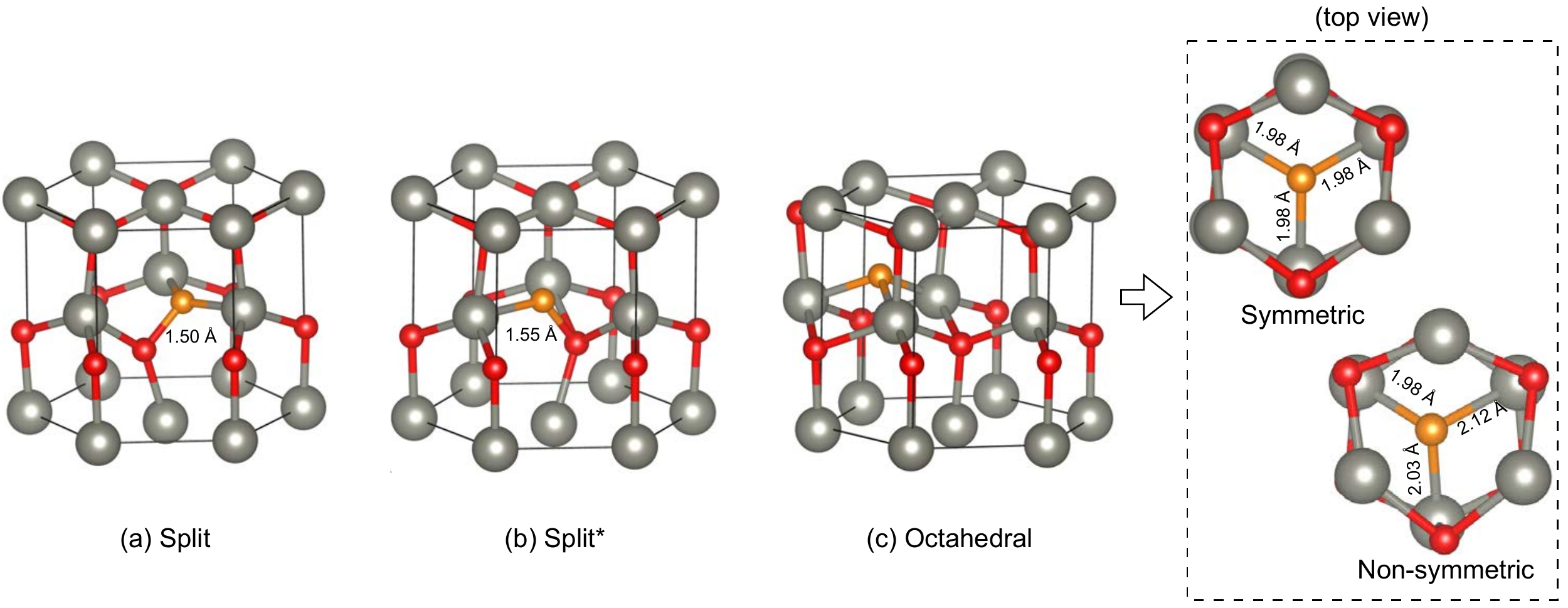}
\caption{\label{fig:7} Oxygen interstitial structures (in orange) in ZnO: (a) split, (b) split*, and (c) octahedral geometry. Octahedral oxygen interstitial has symmetric and non-symmetric configuration as shown in projection along [001] Z-axis in the top view.}
\end{figure*}

Vacancy structures are analyzed in terms of distances between the neighboring 4 silicon atoms to the vacancy site. It has been reported\cite{Watkins1997} that silicon vacancy undergo structure reconstruction in different charge states, and form a Negative-U system.\cite{Baraff1979, Watkins1997} The negative-U behavior implies an energy lowering structural distortion by the presence of a second electron, such that the energy gain more than compensates the \textit{e-e} repulsive energy cost.\cite{Baraff1979} For the spin-polarized calculations, vacancy in the neutral and 2- charge state relaxes to higher energy configurations, with C$_{2v}$ and D$_{2d}$ point group symmetry, respectively. The inability of spin-polarized calculations to reproduce lower energy point group symmetry of charged vacancies has also been reported in previous LDA\cite{Corsetti2011} and GGA\cite{Spiewak2013} calculations. We observe (2+/1+) and (1+/0) charge transitions for silicon vacancy, instead of the direct (2+/0) charge transition because the computed neutral Si vacancy is in the higher energy configuration compared to its lower energy D$_{2d}$ configuration. However, using spin-polarized HSE calculations on DFT structures, we observe direct (2+/0) charge transition.\par

Among interstitials, tetrahedral structure is most stable in 2+ charge state, which then transition (-0.27 eV below CBM) to the split structure, which is the most stable configuration for 0 and 1- charge states. The computed defect formation energies and charge transition levels (Fig. \ref{fig:6}) for silicon vacancies and interstitials are in good agreement with the reported GGA\cite{Rinke2009, Corsetti2011, Spiewak2013} and HSE calculations.\cite{Weber2013, Spiewak2013} However, their are noticeable difference in the charge defect transition levels between GGA and GGA+GW, mainly due to the band edge positions predicted by the self-consistent GW calculation.\cite{Chen2014a} Similar differences in charge transition levels between LDA and LDA+G$_{0}$W$_{0}$ in calculations on silicon interstitials has been reported by Rinke et al.\cite{Rinke2009}

\subsection{\label{sec:IIIB} Zinc Oxide}

ZnO is a direct band gap semiconductor and occurs in the ground state wurtzite crystal structure (space group P6$_{\mathrm{3}}$mc, 186), with two lattice parameters, $a$  and $c$, in the ratio of $c/a = 1.63$. The calculated lattice constants and band gap for the wurtzite ZnO are in good agreement with the known experimental measurements, as summarized in table \ref{tab:1}. Comprehensive studies of intrinsic vacancy and interstitial structures\cite{Janotti2007, Huang2009} in ZnO has been done in the past. In the following discussion we analyze the defect structures predicted using the automated defect framework and compare our results with the existing literature.\par

\begin{figure}[b]
\includegraphics[width=1.\linewidth]{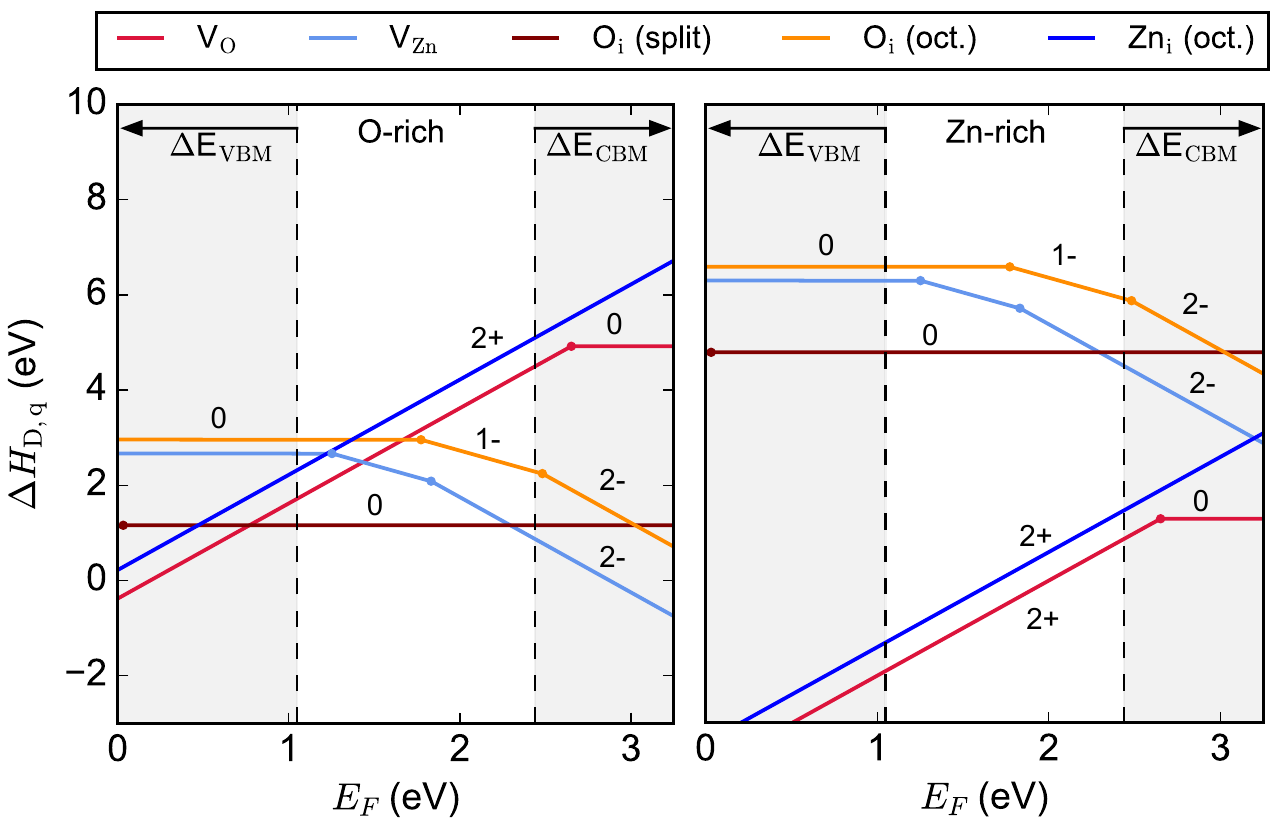}
\caption{\label{fig:8} Defect formation energy as function of the Fermi energy for intrinsic vacancies and interstitials in ZnO using GGA+U with band edge shift computed from GW calculations.}
\end{figure}

\begin{figure*}
\includegraphics[width=0.7\textwidth]{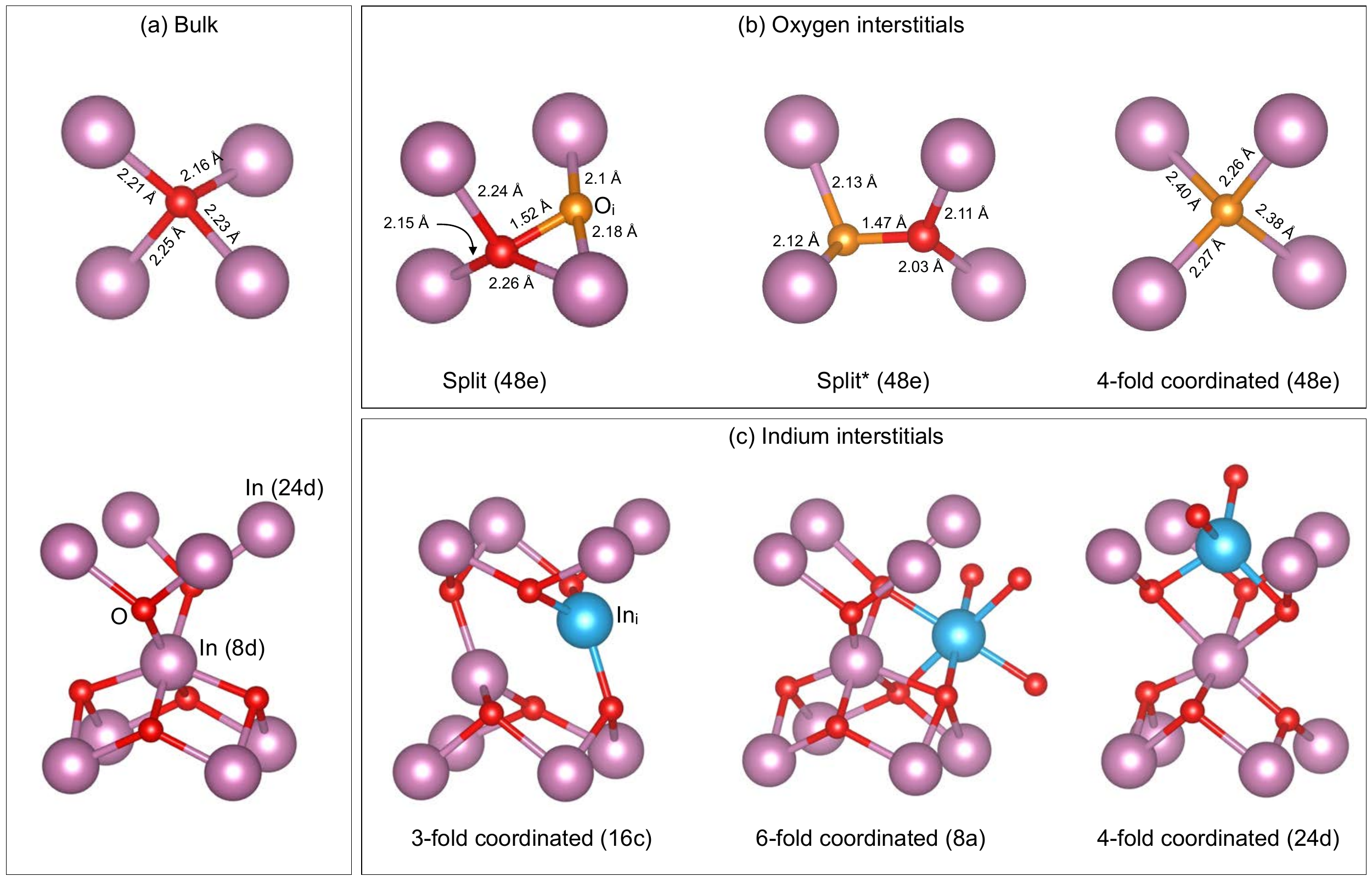}
\caption{\label{fig:9} (a) Oxygen site at 48e (top) and Indium sites at 8b and 24d (bottom) Wyckoff positions in bulk In$_{2}$O$_{3}$, (b) oxygen interstitials (orange) in split(48e) , split*(48e) and 4-fold coordinated (48e), (c) Indium interstitial (blue) in 3-fold (16c), 6-fold (8a), and  4-fold coordinated (24d) configurations.}
\end{figure*}

Figure \ref{fig:7} shows the obtained three distinct oxygen interstitial structures out of the starting 17 possibilities for the interstitial sites. Among oxygen interstitials, split interstitial (Fig. \ref{fig:7}(a)) has the lowest formation energy. Split interstitial has a metastable configuration (Fig. \ref{fig:7}(b), referred here, and in literature as split*\cite{Erhart2006b, Janotti2007}) which is 0.21 eV higher in energy than the stable split configuration and relaxes to stable configuration on further relaxation. Octahedral oxygen interstitial lies in the empty channel along [001] Z-axis inside the six member ring formed by O-Zn atoms, Fig. \ref{fig:7}(c), and is about 1.8 eV higher in energy than the neutral split interstitial. We find two configurations of octahedral oxygen interstitial as shown in the projection along [001] Z-axis (dashed box) in Fig. \ref{fig:7}. The symmetric configuration is 0.3 eV higher in energy than the non-symmetric configuration. The symmetric octahedral is only stable in the neutral charge state and relaxes to non-symmetric configuration for positive and negative charge states. The low energy non-symmetric octahedral configuration has been reported in a previous DFT study,\cite{Huang2009} investigating migration path of oxygen interstitials along [001] direction. With our method we directly find the non-symmetric configuration as the lowest energy octahedral structure.\par

Zinc interstitial is stable in the octahedral configuration with interstitial atom symmetrically placed at the center of the empty channel along [001] Z-axis, similar to the symmetric octahedral oxygen interstitial. In the relaxed geometry the Zn$_{i}$-O distance is 2.05 \AA, and Zn$_{i}$-Zn distance is 2.45\AA. Among interstitials split oxygen interstitial is stable in neutral charge state for the whole range of Fermi Energy (Fig. \ref{fig:8}). Oxygen interstitials at the octahedral site act as deep acceptors, and have relatively high formation energies compared to Zinc vacancies. Zinc interstitials act as shallow donors, with 2+ charge as the most stable charge state (Fig. \ref{fig:8}). But with formation energies as high as 2.6 eV at CBM, even under Zn rich conditions are unlikely to form in substantial concentration.\par 

Both oxygen and zinc in ZnO occupy the 2b Wyckoff position, with 4-fold coordinated tetrahedral geometry. Oxygen vacancy in 2+ charge state shows relatively large outward relaxation of the neighboring Zn atoms, as reported in previous DFT calculations.\cite{Janotti2007} Oxygen vacancy shows transition from 2+ to 0 charge state, at Fermi energy -0.45 eV below the CBM (Fig. \ref{fig:8}), confirming the reported Negative-U character.\cite{Janotti2007, Lany2007, Oba2008} Oxygen vacancy act a deep donor, with fully occupied neutral defect state inside the band gap. Zinc vacancy has partially occupied defect states in the band gap, and act as deep acceptor with (0/1-) and (1-/2-) transition level occur at 1.16, and 1.58 eV, respectively above the VBM.\par

Overall, our approach confirms the known interstitial and vacancy structures in ZnO, and provide a clear picture of the defect energetics and electronic structure consistent with the previous defect calculations. Our next step forward is to investigate the automated point defect framework against In$_{2}$O$_{3}$,  a relatively complex crystal structure containing 40 atoms in the primitive unit cell.


\subsection{\label{sec:IIIA} Indium Oxide}

In$_{2}$O$_{3}$ is a direct band gap semiconductor which is widely used as a transparent conducting oxide. Intrinsic defects in In$_{2}$O$_{3}$ have received relatively moderate attention both experimentally\cite{DeWit1977, DeWit1977a} and theoretically\cite{Lany2007, Agoston2009a, Liu2014} compared to silicon and ZnO. It crystalizes in ground state cubic bixbyite structure (space group Ia-3, 206) with indium (Wyckoff positions 8b and 24d) and oxygen (48e) lattice sites in the bulk structure as shown in Fig. \ref{fig:9}(a).\par

Figure \ref{fig:9} displays the oxygen and indium interstitials structures realized using the automated defect framework. We observe three distinct structures for oxygen interstitials (among the initial set of 22 possible candidates), all occupying the general 48e (x, y, z) Wyckoff position in the relaxed structure. Split oxygen interstitial (Fig. \ref{fig:9}(b)) is the lowest energy configuration. We find a new split interstitial configuration (referred as split* in Fig. \ref{fig:9}(b)) which is stable in the neutral charge state and is about 0.67 eV higher than the lowest energy split interstitial. However split* configuration is unstable in positive charge states and relaxes to the split geometry. Oxygen interstitial bonded to 4 neighboring indium atoms (referred as 4-fold coordinated in Fig. \ref{fig:9}(b)) is the highest energy configuration, with energy of about 1.0 eV higher than the split configuration. To our knowledge only the split and 4-fold coordinated oxygen interstitial configuration has been reported in literature.\cite{Lany2007} This could be due to the fact that, first, such an exhaustive method to theoretically search interstitials has not been adopted for In$_{2}$O$_{3}$, and second, often only the un-occupied Wyckoff positions (16c and 8a) are considered to investigate interstitials in In$_{2}$O$_{3}$.\cite{Lany2007}\par

\begin{figure}
\includegraphics[width=1.\linewidth]{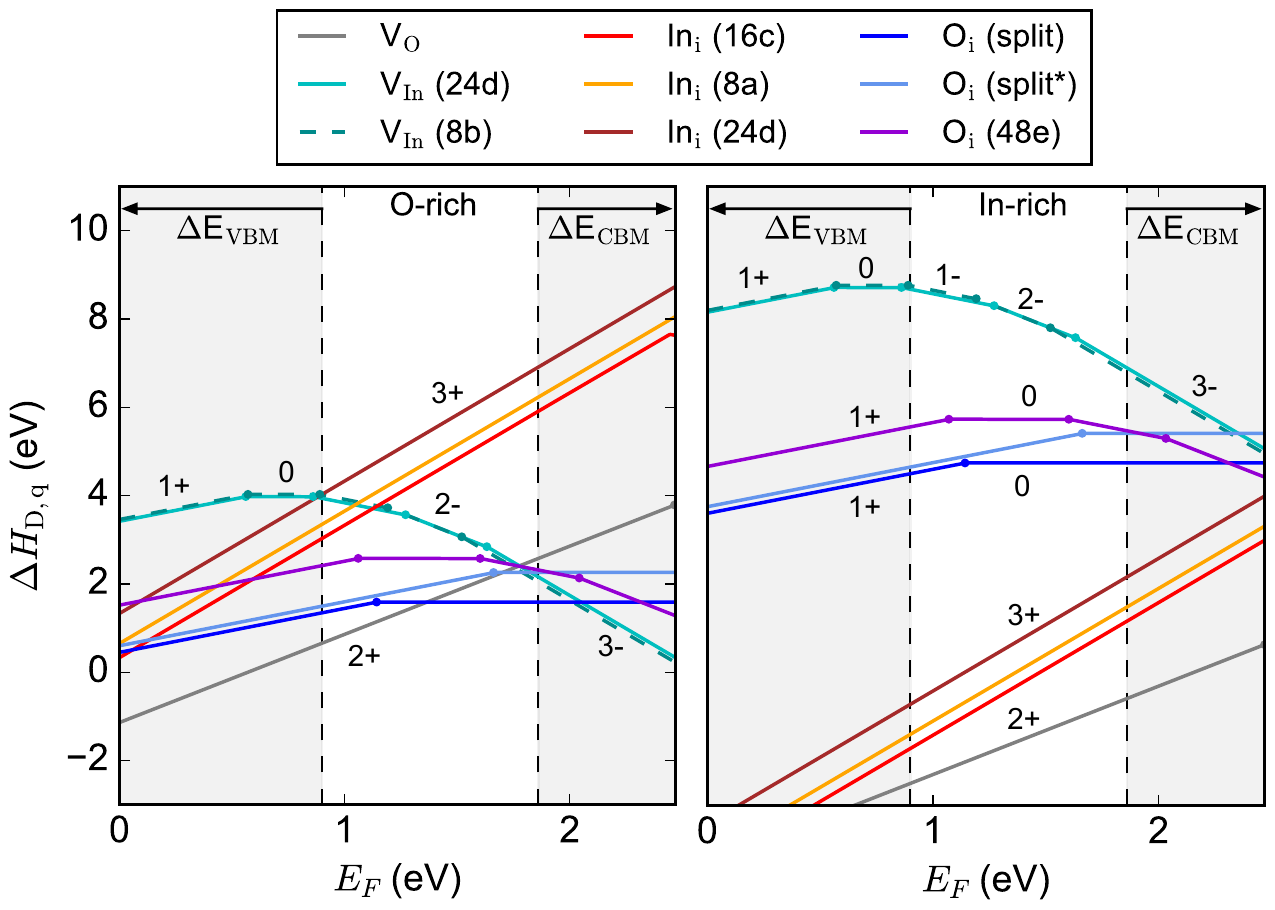}
\caption{\label{fig:10} Defect formation energy as function of the Fermi energy for intrinsic vacancies and in In$_{2}$O$_{3}$ using GGA with band edge shifts computed from GW calculation.}
\end{figure}

For indium interstitials we also find three distinct configurations (Fig. \ref{fig:9}(c)). Lowest energy configuration correspond to indium atom occupying the 16c (0.17, 0.17, 0.17) Wyckoff position, which is 3-fold coordinated to neighboring oxygen atoms (bond length 2.15 \AA). Interstitial atom displaces the indium atom originally at the lattice site 8b (with 6-fold coordination) to a similar 16c (0.83, 0.17, 0.34) Wyckoff position with 3-fold coordination. Indium interstitial at 8a (0.5, 0.0, 0.5) Wyckoff position is symmetrically placed in the empty channel between host indium and oxygen atoms along [110] type direction. It is bonded to 6 neighboring oxygen atoms at bond length of 2.24 \AA. In neutral charge state its energy is about 0.3 eV higher than 16c configuration. Third indium interstitial configuration occupy the 24d (0.75, 0.25, 0.50) Wyckoff position, and is also placed in the empty channel along [110] type direction. It is bonded to 4 neighboring oxygen atom, two of which are at bond length 2.05 \AA, and the other two at 2.18 \AA. 24d configuration is highest in energy, with energy 0.83 eV than the 16c configuration in the neutral charge state. To our knowledge, 24d configuration for indium interstitial has never been considered previously, and though it is high in energy, we believe its existence is relevant and crucial because of the entropy at growth temperatures.\par

In the context of electronic structure, oxygen interstitials has defect states deep inside the band gap, and so as for the indium vacancies (Fig. \ref{fig:10}). 
Indium vacancies have two distinct configurations 8b (0.75, 0.25, 0.25) and 24d (0.75, 0.5, 0.0) both 6-fold coordinated to the neighboring oxygen atoms. Indium interstitials in all structural configurations act as shallow donors, with defect states formed as resonance states above the CBM. Indium interstitial occur in 3+ charge state for the Fermi energy in majority of the band gap, with charge transition levels occur almost at the CBM (Fig. \ref{fig:10}). We observe shallow donor type defect states for oxygen vacancy in DFT with 2+ charge state being the most stable within the explored range of the Fermi energy. As discussed previously,\cite{Lany2011} defect states in DFT can hybridize strongly with the band edges, and requires self-consistent band gap corrected method such as hybrid functional and defect GW to accurately determine oxygen vacancy charge transition levels. Overall, our results are consistent with previous DFT calculations\cite{Lany2007, Liu2014} in In$_{2}$O$_{3}$ and demonstrate the potential of the employed automated point defect framework to discover interstitials structures in complex crystal structures.

\section{\label{sec:IV}Summary and future outlook}

We have developed an efficient and extensively validated framework to automate point defect calculations. We applied the framework to Si, ZnO and In$_2$O$_3$, and recovered the known intrinsic defect structures as well as their electronic structure properties. Our results demonstrate that the automated defect framework can not only be employed to discover interstitials in complex crystal structures such as In$_{2}$O$_{3}$, but also predict accurate defect formation energy of point defects using the implemented finite-size correction schemes. The package is being continuously developed and is hosted on GitHub at \href{https://github.com/pylada/pylada-defects}{https://github.com/pylada/pylada-defects}. We believe an automated point defect analysis framework like this will accelerate structure-property prediction by bringing detailed defect understanding to the forefront, and will contribute to more strategic efforts towards tuning the device performance.

\begin{acknowledgments}

We thank Rachel Kurchin for helpful discussions. A. Goyal and V. Stevanovic are funded by the National Science Foundation (NSF) partially under grants DMR-1309980 and CBET-1605495. P. Gorai is funded by the NSF DMR program, grant no. 1334713. H. Peng and S. Lany acknowledges support as part of the Center for the Next Generation of Materials by Design, an Energy Frontier Research Center (EFRC) funded by U.S. Department of Energy (DOE), Office of Science, Basic Energy Sciences. This research used computational resources sponsored by the DOE Office of Energy Efficiency and Renewable Energy and located at the National Renewable Energy Laboratory (NREL).

\end{acknowledgments}

\bibliography{manuscript}

\end{document}